\documentclass[twoside,leqno,twocolumn]{article}

\usepackage[letterpaper]{geometry}

\usepackage{ltexpprt}

\usepackage{algorithm}
\usepackage[algo2e,ruled,vlined,linesnumbered]{algorithm2e}
\usepackage{booktabs}
\usepackage{cite}
\usepackage{xspace}
\usepackage{color}
\usepackage{xcolor}
\usepackage{url}
\usepackage{hyperref}
\usepackage{amsmath}
\usepackage{todonotes}
\usepackage{cleveref}
\usepackage{placeins}

\newif\ifTR
\TRtrue
\usepackage{amsfonts}

\newcommand{\set}[1]{\left\{ #1\right\}}

\newcommand{\sodass}{\,:\,}
\newcommand{\setGilt}[2]{\left\{ #1\sodass #2\right\}}

\newcommand{\realrange}[2]{\left[#1, #2\right]}

\newcommand{\unitrange}[2]{\realrange{0}{1}}

\newcommand{\Oh}[1]{\mathcal{O}\!\left( #1\right)}

\newcommand{\discussionsize}{\small}

\newcommand{\frage}[1]{}

\newsavebox{\codeparam}

{\end{disscodepos}}

\newcommand{\Is}       {:=}

\newdimen\endofsize\endofsize=0.5em
\def\endofbeweis{~\quad\hglue\hsize minus\hsize
                 \hbox{\vrule height \endofsize width
\endofsize}\par}

\usepackage{numprint}
\usepackage{algorithmic}
\usepackage{graphicx}
\def\mate{\mathrm{mate}}

\newcommand{\ie}{i.e.,\xspace}
\newcommand{\etal}{et~al.\xspace}
\newcommand{\eg}{e.g.}
\newcommand{\algo}[1]{\textsc{#1}}
\newcommand{\dynrand}{\algo{DynMWMRandom}\xspace}
\newcommand{\dynlevel}{\algo{DynMWMLevel}\xspace}
\newcommand{\dynlevelR}{\algo{DynMWMLevelR}\xspace}
\newcommand{\dynlevelB}{\algo{DynMWMLevelB}\xspace}
\renewcommand{\Oh}{\mathcal{O}}

\newtheorem{claim}{Claim}[section]

\ifTR
\else \fi{}
\newcommand{\mytitle}{Fully-dynamic Weighted Matching Approximation in Practice}
\begin{document}
\title{\mytitle}

\pagestyle{headings}

\newcommand*\samethanks[1][\value{footnote}]{\footnotemark[#1]}

\ifTR
\author{Eugenio Angriman\thanks{Humboldt-Universität zu Berlin, Berlin, Germany} \and Henning Meyerhenke\samethanks[1] \and Christian Schulz\thanks{Heidelberg University, Heidelberg, Germany} \and Bora Uçar\thanks{CNRS and LIP, ENS Lyon, France}}
\else
\author{}
\fi 
\date{}

\maketitle

\begin{abstract}
Finding large or heavy matchings in graphs is a ubiquitous combinatorial optimization problem.
In this paper, we engineer the first non-trivial implementations for approximating
the dynamic weighted matching problem.
Our first algorithm is based on random walks/paths combined with dynamic programming.
The second algorithm has been introduced by Stubbs and Williams without an implementation.
Roughly speaking, their algorithm uses dynamic unweighted matching algorithms as a subroutine
(within a multilevel approach); this allows us to use previous work on dynamic unweighted 
matching algorithms as a black box in order to obtain a fully-dynamic weighted matching algorithm. 
We empirically study the algorithms on an extensive set of dynamic instances and compare them with 
optimal weighted matchings. Our experiments show that the random walk algorithm
typically fares much better than Stubbs/Williams (regarding the time/quality tradeoff), and its
results are often not far from the optimum.

\end{abstract}
\thispagestyle{empty}
\section{Introduction}
\label{sec:intro}
%
A matching in a graph is a set of pairwise vertex-disjoint edges.
Alternatively, a matching can be seen as a subgraph (restricted to its edges) with degree at most $1$. 
A matching is \emph{maximal} if no edges can be added to it without
violating the matching property that no two matching edges share a common vertex.
A matching of a graph $G$ is \emph{maximum}, in turn, if there exists
no matching in $G$ with higher cardinality.
Computing (such) matchings in a graph is a ubiquitous combinatorial 
problem that appears in countless applications~\cite{Halappanavar09algorithms}.
Two popular optimization problems in this context are 
(i) the maximum cardinality matching (MCM) problem, which seeks
a matching with maximum cardinality,
and (ii) the maximum weighted matching (MWM) problem, \ie to find a 
matching (in a weighted graph) whose total edge weight is maximum.
Micali and Vazirani~\cite{DBLP:conf/focs/MicaliV80} solve MCM in
$\Oh(m\sqrt{n})$ time, Mucha and Sankowski~\cite{DBLP:journals/algorithmica/MuchaS06}
in $\Oh(n^\omega)$, where $\omega < 2.373$ is the matrix
multiplication exponent~\cite{DBLP:journals/corr/abs-2010-05846}, $m$ is the number of edges,
and $n$ is the number of vertices.
Concerning MWM, the best known algorithm is by Galil \etal~\cite{DBLP:journals/csur/Galil86}, it takes
$\Oh(mn\log n)$ time.
For integral edge weights up to $W$, 
Gabow and Tarjan~\cite{DBLP:journals/jacm/GabowT91} proposed an
$\Oh(m\sqrt{n}\log{nW})$ time algorithm.
Sankowski's algorithm~\cite{DBLP:journals/tcs/Sankowski09}, in turn, requires
$\tilde{\mathcal{O}}(Wn^\omega)$ time, where
$\tilde{\mathcal{O}}$ hides a polylogarithmic factor. 

Real-world graphs occurring in numerous scientific and commercial
applications are not only massive but can also change over
time~\cite{DBLP:conf/focs/MehtaSVV05}. For example, edges can be inserted,
deleted, or their weight can be updated 
-- think of a road network where edge weights represent the traffic
density, the constantly evolving web graph, or the virtual friendships
between the users of a social network.
Hence, recomputing a (weighted) matching from scratch each time the graph changes
is expensive. With many frequent changes, it can even be prohibitive,
even if we are using a polynomial-time algorithm.
Further, when dealing with large graphs,
computing an optimal matching can be too time-consuming and thus one often resorts
to approximation to trade running time with solution quality.
In recent years, several fully-dynamic algorithms for both exact and
approximate MCM~\cite{IvkovicL93,OnakRubinfeld10,
BaswanaGS15,Solomon16,BhattacharyaHI18,Bhattacharya2016,
BhattacharyaCH17,DBLP:conf/focs/GuptaP13,NeimanS16,Bernstein2016a,CharikarS18,ArarCCSW18,
DBLP:conf/soda/0001LSSS19,DBLP:journals/jea/BarenboimM19,kashyop2020105982,Sankowski07}
and MWM~\cite{DBLP:conf/fsttcs/AnandBGS12,GuptaP13,DBLP:conf/innovations/StubbsW17}
have been proposed.
These algorithms exploit previously computed information about the matching to
handle graph updates more efficiently than a static recomputation.
Yet, limited effort has been invested to actually implement these algorithms
in executable code and test their performance on real-world instances.
Only recently, Henzinger \etal\cite{DBLP:conf/esa/Henzinger0P020} engineered several
fully-dynamic algorithms for MCM and investigated their practical performance on large real-world
graphs. Similar results have not been produced yet for fully-dynamic MWM algorithms 
(neither exact nor approximate) and therefore their practical performance is still unknown.

\paragraph*{Contributions.} 
In this paper we focus on the fully-dynamic MWM problem
and explore the gap between theoretical results and practical performance for two algorithms.
The first one, inspired by Maue and Sanders~\cite{MauSan07}, combines random walks and dynamic
programming to compute augmenting paths in the graph; if the random walks are of appropriate length
and repeated sufficiently often, the algorithm maintains a $(1 + \epsilon)$-approximation
in $\Oh(\epsilon^{-1}\Delta^{2/\epsilon+3} \log n)$ time per update, where $\Delta$ is the maximum node degree observed throughout the algorithm.
The second algorithm, introduced by Stubbs and Williams~\cite{DBLP:conf/innovations/StubbsW17},
uses existing fully-dynamic $\alpha$-approximation algorithms for MCM as subroutines to maintain
a $2\alpha(1 + \epsilon)$-approximation for MWM in fully-dynamic graphs.
We provide the first implementations of the aforementioned algorithms,
analyze their performance in systematic experiments on an extensive set
of dynamic instances, and compare the quality of the computed matchings 
against the optimum. The best algorithm is very often less than 10\% away from the optimum
and at the same time very fast. 

\newcommand{\EM}{${\cal E}_M$\xspace}
\newcommand{\Em}{{\cal E}_M}

\section{Preliminaries}

\paragraph*{Basic Concepts.} 
Let  $G=(V,E)$ be a simple \emph{undirected graph} with edge weights $\omega: E \rightarrow \mathbb{R}_{> 0}$.
  We extend $\omega$ to sets, i.e., $\omega(E') := \sum_{e
  \in E'} \omega(e)$. 
 We set $n = |V|$, and $m = |E|$;
$N(v)\Is \setGilt{u}{\set{v,u}\in E}$ denotes the \emph{neighbors} of $v$.
The (unweighted) degree of a vertex $v$ is $d(v):=|N(v)|$.
A matching $\mathcal{M} \subset E$ in a graph is a set of edges without common vertices.
The \emph{cardinality} or \emph{size} of a matching is simply the cardinality of the edge subset $\mathcal{M}$.
We call a matching \emph{maximal} if there is no edge in $E$ that can be added to $\mathcal{M}$. 
A \textit{maximum cardinality matching} $\mathcal{M}_{\text{opt}}$ is a matching that contains the largest possible
number of edges of all matchings.
A \textit{maximum weight matching} $\mathcal{M'}_{\text{opt}}$ is a matching that maximizes $\omega(\mathcal{M'}_{\text{opt}})$ among all possible matchings.
An $\alpha$-\emph{approximate} maximum (weight) matching is a matching that has weight at least $\frac{\omega(\mathcal{M}_{\text{opt}})}{\alpha}$.
A vertex is called \emph{free} or \emph{unmatched} if it is not incident to an edge of the matching. Otherwise, we call it \emph{matched}.
For a matched vertex $u$ with $\{u,v\} \in \mathcal{M}$, we call vertex $v$ the \emph{mate} of $u$, which we denote as $\mate(u)=v$. 
For an unmatched vertex $u$, we define $\mate(u)=\bot$. 
An \emph{augmenting path} is defined as a cycle-free path in the graph $G$ that starts and ends on a \emph{free} vertex and where edges from $\mathcal{M}$ alternate with edges from $E \setminus \mathcal{M}$. The \emph{trivial augmenting path} is a single edge with both endpoints free. Throughout this paper, we call such an edge a \emph{free} edge. If we take an augmenting path and resolve it by matching every unmatched edge and unmatching every matched edge, we increase the cardinality of the matching  by one.
In the maximum cardinality case, any matching without \emph{augmenting paths} is a maximum matching~\cite{berge57} and  any matching with no augmenting paths of length at most $2k-3$ is a $(k/(k-1))$-approximate~maximum~matching~\cite{hopkarp71}. 
A {\textit{weight-augmenting path}} $\mathcal{P}$ with respect to a matching $\mathcal{M}$ is an alternating path whose free edges are heavier than
its edges in $M$: $w(\mathcal{M}\oplus \mathcal{P})> w(\mathcal{M})$,
where $\oplus$ denotes the symmetric difference.
A weight-augmenting path with $k$ edges outside $\mathcal{M}$ is called {\textit{weight-augmenting $k$-path}}.
Note that a weight-augmenting $k$-path can have $2k-1, 2k$, or $2k+1$ edges, 
whereas in the unweighted case an augmenting path with $k$ edges outside of a given matching has exactly $2k-1$ edges.

Our focus in this paper are \emph{fully-dynamic graphs}, where the number of vertices is fixed, but edges can be added and removed.\footnote{The sole focus on dynamic edges is no major limitation:
one could simply insert enough singleton vertices into the original graph
to mimic vertex insertions. Only new edges inserted over time would make 
the new vertices eligible as matching partners.}
All the algorithms evaluated can handle edge insertions as well as edge deletions.
In the following, $\Delta$ denotes the maximum degree that can be found in any state of the dynamic graph.


\paragraph*{Related Work.} 
\label{sec:relatedwork}
Dynamic algorithms are a widely researched topic. We refer the reader to the recent survey by Hanauer \etal \cite{DBLP:journals/corr/abs-2102-11169} for most material related to theoretical and practical dynamic algorithms.
Many of the numerous matching applications require 
matchings with certain properties, like maximal (no edge can
be added to $\mathcal{M}$ without violating the matching property) or maximum
cardinality matchings. 
Edmonds's blossom algorithm~\cite{edmonds1965paths} computes
a maximum cardinality matching in a static graph in time $\Oh(mn^2)$. This result was later improved to $\Oh(m \sqrt{n})$ by Micali and Vazirani \cite{DBLP:conf/focs/MicaliV80}. Some recent algorithms use simple data reductions rules~\cite{DBLP:conf/esa/KorenweinNNZ18} or shrink-trees instead of blossoms~\cite{DBLP:conf/alenex/DroschinskyMT20} to speed up computations in static graphs. 
In practice, these algorithms can still be time-consuming for many applications involving large graphs. 
Hence, several practical approximation algorithms with nearly-linear running time exist such as the local max algorithm \cite{DBLP:conf/europar/BirnOSSS13}, the path growing algorithm~\cite{DH03a}, global paths~\cite{MauSan07},
and suitor~\cite{DBLP:conf/ipps/ManneH14}. 
As this paper focuses on dynamic graphs, we refer the reader to the quite extensive related work section of~\cite{DBLP:conf/alenex/DroschinskyMT20} for more recent static matching algorithms.

In the dynamic setting, the maximum matching problem has been prominently studied ensuring $\alpha$-approximate guarantees. 
A major exception is the algorithm by van den Brand \etal~\cite{DBLP:conf/focs/BrandNS19}, which maintains the \emph{exact} size of a maximum matching in $\Oh(n^{1.407})$ update time.
One can trivially maintain a maximal ($2$-approximate) matching in $\Oh(n)$ update time by resolving all trivial augmenting paths of length one. 
Ivkovi\'c and Llyod \cite{IvkovicL93} designed the first fully-dynamic algorithm to improve this bound to $\Oh((n+m)^{\sqrt{2}/2})$ update time. 
Later, Onak and Rubinfeld~\cite{OnakRubinfeld10} presented a randomized algorithm for maintaining an $\Oh(1)$-approximate matching in a dynamic graph that takes $\Oh(\log^2 n)$ expected amortized time for each edge update. 
This result led to a flurry of results in this area. 
Baswana, Gupta and Sen~\cite{BaswanaGS15} improved the approximation ratio of \cite{OnakRubinfeld10} from $\Oh(1)$ to $2$ and the amortized update time to $\Oh(\log n)$. 
Further, Solomon~\cite{Solomon16} improved the update time of~\cite{BaswanaGS15} from amortized $\Oh(\log n)$ to {\em constant}. 
However, the first deterministic data structure improving~\cite{IvkovicL93} was given by  Bhattacharya et al.~\cite{BhattacharyaHI18}; it maintains a $(3+\epsilon)$ approximate matching in $\tilde{O}(\min(\sqrt{n},m^{1/3}/\epsilon^2))$ amortized update time, which was further improved to $(2+\epsilon)$ requiring $\Oh(\log n/\epsilon^2)$ update time by Bhattacharya et al.~\cite{Bhattacharya2016}. 
Recently, Bhattacharya et al.~\cite{BhattacharyaCH17} achieved the first $\Oh(1)$ amortized update time for a deterministic algorithm but for a weaker approximation guarantee of $\Oh(1)$. 
For worst-case bounds, the best results are by (i) Gupta and Peng~\cite{GuptaP13} requiring $\Oh(\sqrt{m}/\epsilon)$ update time for a $(1+\epsilon)$-approximation, (ii) Neiman and Solomon~\cite{NeimanS16} requiring $\Oh(\sqrt{m})$ update time for a $(3/2)$-approximation, and (iii) Bernstein and Stein~\cite{Bernstein2016a} requiring $\Oh(m^{1/4}/\epsilon^{2.5})$ update time for a $(3/2+\epsilon)$-approximation. 
Recently, Charikar and Solomon~\cite{CharikarS18} as well as Arar \etal~\cite{ArarCCSW18} (using \cite{Bhattacharya2017b}) presented independently the first algorithms requiring $\Oh(poly\log n)$ worst-case update time while maintaining a $(2+\epsilon)$-approximation.
Recently, Grandoni~\etal\cite{DBLP:conf/soda/0001LSSS19} gave an incremental matching algorithm that achieves a $(1+\epsilon)$-approximate matching in constant deterministic amortized time. 
Barenboim and Maimon \cite{DBLP:journals/jea/BarenboimM19} present an algorithm that has $\tilde{\mathcal{O}}(\sqrt{n})$ update time for graphs with constant neighborhood independence. 
Kashyop and Narayanaswamy~\cite{kashyop2020105982} give a conditional lower bound for the update time, which is sublinear in the number of edges for two subclasses of fully-dynamic algorithms, namely lax and eager algorithms.

Despite this variety of different algorithms, to the best of our knowledge,
very limited efforts have been made so far to engineer these dynamic algorithms
and to evaluate them on real-world instances.
Henzinger \etal~\cite{DBLP:conf/esa/Henzinger0P020} have started to evaluate algorithms for the dynamic maximum cardinality matching problem in practice.
To this end, the authors engineer several dynamic maximal matching algorithms as well as an algorithm that is able to maintain the maximum matching. 
The algorithms implemented in their work are Baswana, Gupta and Sen \cite{BaswanaGS15}, which performs edge updates in  $\Oh(\sqrt{n})$ time and maintains a 2-approximate maximum matching, the algorithm of Neiman and Solomon~\cite{NeimanS16}, which takes $\Oh(\sqrt{m})$ time to maintain a $(3/2)$-approximate maximum matching, as well as two \emph{novel} dynamic algorithms: a random walk-based algorithm as well as a dynamic algorithm that searches for augmenting paths using a (depth bounded) blossom algorithm. 
Experiments indicate that maintaining optimum matchings can be done much more efficiently than the naive algorithm that recomputes maximum matchings from scratch (more than an order of magnitude faster). 
Second, all inexact dynamic algorithms that have been considered in that work are able to maintain near-optimum matchings in practice while being multiple orders of magnitudes faster than the naive optimum dynamic algorithm. The study concludes that in practice an extended random walk-based algorithms should be the method of choice. 

For the \emph{weighted} dynamic matching problem, Anand \etal \cite{DBLP:conf/fsttcs/AnandBGS12} propose an algorithm that can maintain an $4.911$-approximate dynamic maximum weight matching that runs in amortized $\Oh(\log n \log D)$ time, where $D$ is the ratio between the highest and the lowest edge weight.
Gupta and Peng \cite{DBLP:conf/focs/GuptaP13} maintain a $(1+\epsilon)$-approximation under edge insertions/deletions that runs in time $\Oh(\sqrt{m}\epsilon^{-2-\Oh(1/\epsilon)}\log N)$ time per update. Their result is based on
several ingredients: (i) rerunning a static algorithm from time to time, 
(ii) a trimming routine that trims the graph to a smaller equivalent graph whenever possible and 
(iii) in the weighted case, a partition of the edges (according to their weights) into geometrically shrinking intervals.

Stubbs and Williams \cite{DBLP:conf/innovations/StubbsW17} present metatheorems for dynamic weighted matching.
They reduce the dynamic maximum weight matching problem to the dynamic maximum cardinality matching problem in which the graph is unweighted. The authors prove that using this reduction, if there is an $\alpha$-approximation for maximum cardinality matching with update time $T$, then there is also a $2\alpha(1+\epsilon)$-approximation for maximum weight matching with update time $\Oh(\frac{T}{\epsilon^2}\log^{2} C)$. Their basic idea is an extension/improvement of the algorithm of Crouch and Stubbs \cite{DBLP:conf/approx/CrouchS14} who tackled the problem in the streaming model. 
We go into more detail in Section~\ref{sec:stubbwilliams}. None of these algorithms have been implemented.

\section{Algorithm \dynrand: Random Walks + Dynamic Programming}
\label{sec:algorithms}
Random walks have already been successfully employed for dynamic maximum cardinality matching by Henzinger \etal \cite{DBLP:conf/esa/Henzinger0P020}; however, in their current form they cannot be used for dynamic maximum weight matching.
The main idea of our first dynamic algorithm for the weighted dynamic 
matching problem is to find random augmenting paths in the graph.
To this end, our algorithm uses a random walk process in the graph
and then uses dynamic programming on the computed random path $\mathcal{P}$
to compute the best possible matching on $\mathcal{P}$.\footnote{One could rather speak solely of ``random paths'' instead of ``random walks''. 
We decided to keep the ``walk'' term, because the underlying process
is that of a (cycle-free) random walk.}
We continue by explaining these concepts in our context and how we handle edge insertions and deletions.

\subsection{Random Walks For Augmenting Paths.}
Our first dynamic algorithm is called \dynrand and constructs a random (cycle-free) path $\mathcal{P}$ as follows:
initially, all nodes are set to be eligible. Whenever an edge is added to the path $\mathcal{P}$, we set the endpoints to be not eligible.
The random walk starts at an arbitrary eligible vertex $u$.
If $u$ is free, the random walker tries to randomly choose an eligible neighbor $w$ of $u$. 
If successful, the corresponding edge $e$ is added to the path $\mathcal{P}$ and the walk is continued at $w$ 
(otherwise the random walk stops and returns $\mathcal{P}$). Moreover, $u$ is set to not eligible.
On the other hand, if $u$ is matched and $\mate(u)$ is eligible, 
we add the edge $\{u, \mate(u)\}$ to $\mathcal{P}$, set $u$ to not eligible and continue the walk at $v:=\mate(u)$.
There the algorithm tries to randomly choose an eligible neighbor $w$ of $v$.
If successful, the random walker adds the corresponding edge to the path, sets $v$ to not eligible and continues the walk at $w$. 
If at any point during the algorithm execution, there is no adjacent eligible vertex, then the algorithm stops and returns the path $\mathcal{P}$.
Note that, by construction, no vertex on the path is adjacent to a matched edge that is not on the path.
We stop the algorithm after $\Oh(\frac{1}{\epsilon})$ steps for a given $\epsilon$.  

The algorithm tries to find a random eligible neighbor by sampling a neighbor $u$ uniformly at random. 
If $u$ is eligible, then we are done.
If $u$ is not eligible, we repeat the sampling step.
We limit, however, the number of unsuccessful repetitions to a constant.
If our algorithm did not find an eligible vertex, we stop and return the current path $\mathcal{P}$. 
Overall, this ensures that picking a random neighbor can be done in constant time. 
Hence, the time to find a random path is $\Oh(\frac{1}{\epsilon})$.
Here, we assume that the eligible state of a node is stored in an array of size $n$ that is used for many successive random walks.
This array can be reset after the random walk has been done in $\Oh(\frac{1}{\epsilon})$ by setting all nodes on the path to be eligible again.
The length of the random walk is a natural parameter of the algorithm that we will investigate in the~experimental evaluation. 

After the path has been computed, we run a dynamic program on the path to compute the optimum weighted matching on the path in time $\Oh(\frac{1}{\epsilon})$.
If the weight of the optimum matching on $\mathcal{P}$ is larger than the weight of the current matching on the path, we replace the current matching on $\mathcal{P}$. 
Note that, due to the way the path has been constructed, this yields a feasible matching on the overall graph.

\paragraph*{Dynamic Programming on Paths.} It is well-known that the weighted matching problem can be solved to optimality on paths using dynamic programming~\cite{MauSan07}. In order to be self-contained, we outline briefly how this can be done. The description of Algorithm~\ref{algo:dynprog} in Appendix~\ref{app:sec:dynprogalgo} follows Maue and Sanders~\cite{MauSan07}. 
Given the path $\mathcal{P}=\langle e_1, \ldots, e_k \rangle$, the main idea of the dynamic programming approach is to scan the edges in the given order. Whenever it scans the next edge $e_i$, there are two cases: either that edge is used in an optimum matching of the subproblem $\langle e_1, \ldots, e_i \rangle$ or not. To figure this out, the algorithm takes the weight of the current edge $e_i$ and adds the weight of the optimum subsolution for the path $\langle e_1, \ldots, e_{i-2}\rangle$. If this is larger than the weight of the optimum subsolution for the path $\langle e_1, \ldots, e_{i-1}\rangle$, then the edge $e_i$ is used in an optimum solution for $\langle e_1, \ldots, e_i\rangle$; otherwise it is not.

\paragraph*{Optimizations.} The random walk can be repeated multiple times, even if it a weight-augmenting path found.
This is because, even if the random walker found an improvement, it can be possible that there is another augmenting path left in the graph. In our experiments, we use $\ell$ walks, where $\ell$ is a tuning parameter. 
However, repeating the algorithm too often can be time-consuming. 
Hence, we optionally use a heuristic to break early if we already performed a couple of unsuccessful random walks: we stop searching for augmenting paths if we made $\beta$ consecutive random walks that were all unsuccessful. In our experiments, we use $\beta=5$; similar choices of parameters should work just as well. We call this the \emph{``stop early'' heuristic}.
We now explain how we perform edge insertions~and~deletions using the augmented random walks introduced above.

\paragraph*{Edge Insertion.} Our algorithm handles edge insertions $\{u,v\}$ as follows:
if both $u$ and $v$ are free, we start a random walk at either $u$ or $v$ (randomly chosen) and make sure that the new edge is always included in the path $\mathcal{P}$.  That means if we start at $u$, then we add the edge $\{u, v\}$ to the path and start the random walk described above at $v$ (and vice versa). If one of the endpoints is matched, say $u$, we add the edges $\{u, \mate(u)\}$ and $\{u,v\}$ to the path $\mathcal{P}$ and continue the walk at $v$. If both endpoints are matched, we add $\{u, \mate(u)\}$, $\{u,v\}$ and $\{v,\mate(v)\}$ to the path and continue the random walk at $\mate(v)$. Note that it is necessary to include matched edges, since the optimum weight matching on the path that we are constructing with the random walk may match the newly inserted edge. Hence, if we did not added matched edges incident to $u$ and $v$, we may end up with an infeasible matching. After the full path is constructed, we run the dynamic program as described above and augment the matching if we found a better one. 

\paragraph*{Edge Deletion.}
\label{sec:rw-edge-out}
In case of the deletion of an edge $\{u,v\}$, we start two subsequent random walks as described above, one at $u$ and one at $v$.
Note that if $u$ or $v$ are matched, then the corresponding matched edges will be part of the respective paths $\mathcal{P}$ that are created by the two random walks.

\subsection{Analysis.}
The algorithm described above, called \dynrand, can maintain a 
$(1+\epsilon)$-approximation if the random walks are of appropriate length and repeated sufficiently often.
To this end, we first show a relation between the existence of a weight-augmenting $k$-path and the quality of a matching. For the proofs, see Appendix \ref{appendixproof}.

\begin{proposition}
\label{thm:kappx}
Let $\mathcal{M}$ be a matching and $k\geq 1$ be the smallest number
such that there is a weight-augmenting $k$-path with respect to $\mathcal{M}$.
Then $w(\mathcal{M})\geq \frac{k-1}{k} \cdot w(\mathcal{M}^*)$, where $\mathcal{M}^*$ is a matching with the maximum weight.
\end{proposition}

\begin{theorem}
\label{thm:dyn-approx-ratio}
Algorithm \dynrand maintains a $(1+\epsilon)$-approximate maximum weight matching with high probability if the length of each path is $\lceil 2/\epsilon+3 \rceil$ and the walks are repeated $\lceil\Delta^{2/\epsilon+3} \log n\rceil$ times. 
\end{theorem}

\section{Algorithm Family \dynlevel based on Results by Crouch, Stubbs and Williams}
\label{sec:stubbwilliams}
The algorithm by Stubbs and Williams~\cite{DBLP:conf/innovations/StubbsW17} can use dynamic \emph{unweighted} matching algorithms as a black box to obtain a weighted dynamic matching algorithm.
We outline the details of the meta-algorithm here, as this is one of the algorithms that we implemented.
Consider a dynamic graph that has integer weights in $[L,N]$. Moreover, let $C=N/L$. The basic idea of the algorithm, due to Crouch and Stubbs \cite{DBLP:conf/approx/CrouchS14}, is to maintain $\Oh(\epsilon^{-1}\log C)$ levels. Each of the levels only contains a subset of the overall set of edges. More precisely, the $i$-th level contains all edges of weight at least $(1+\epsilon)^i$ for $i \in \{\lfloor \log_{1+\epsilon} L\rfloor, \lfloor \log_{1+\epsilon} L\rfloor +1, \ldots, \lfloor \log_{1+\epsilon} N\rfloor\}$. The algorithm then maintains unweighted matchings on each level by employing a dynamic maximum cardinality matching algorithm for each level when an update occurs. The output matching is computed by including all edges from the matching in the highest level, and then by adding matching edges from lower levels in descending order, as long as they are not adjacent to an edge that has been already added. Stubbs and Williams~\cite{DBLP:conf/innovations/StubbsW17} extend the result by Crouch and Stubbs \cite{DBLP:conf/approx/CrouchS14} and show how the greedy merge of the matchings can be updated efficiently.
\begin{theorem}[Stubbs and Williams~\cite{DBLP:conf/innovations/StubbsW17}]
If the matchings on every level are $\alpha$-approximate maximum cardinality matchings, then the output matching of the algorithm is a $2\alpha(1+\epsilon)$-approximate maximum weight matching.
Moreover, the algorithm can be implemented such that the update time is $\Oh(t(m,n) (\log_{1+\epsilon}N/L)^2)$, where $t(m,n)$ is the update time of the dynamic cardinality matching algorithm used.
\end{theorem}
In our implementation, we use the recent dynamic maximum cardinality matching
algorithms due to Henzinger \etal~\cite{DBLP:conf/esa/Henzinger0P020} that fared best in their experiments. 
The first algorithm is a random walk-based algorithm, the second algorithm is a (depth-bounded) blossom algorithm
-- leading to the dynamic counterparts \dynlevelR and \dynlevelB, respectively. The latter also has the option to maintain optimum maximum cardinality matchings.
Both are described next to be self-contained. We refer to~\cite{DBLP:conf/esa/Henzinger0P020} for more details.

\paragraph*{Random Walk-based MCM.} In the update routines of the algorithm
(\emph{insert} and \emph{delete}), the algorithm uses a random walk that works as follows:
The algorithms starts at a free vertex $u$ and chooses a random neighbor $w$ of $u$. 
If the chosen neighbor is free, the  edge $\{u,w\}$ is matched and the random walk stops. 
If $w$ is matched, then we unmatch the edge $\{w, \mate(w)\}$ and match $\{u,w\}$ instead. Note that $u \neq \mate(w)$ since $u$ is free in the beginning and therefore $\mate(u) = \bot$, but $\mate(\mate(w)) = w$ and $w \neq \bot$. Afterwards, the previous mate of $w$ is free. Hence, the random walk is continued at this vertex. 
The random walk performs $\Oh(\frac{1}{\epsilon})$ steps. 
If unsuccessful, the algorithm can perform multiple repetitions to find an augmenting path. Moreover, the most successful version of the algorithm uses \emph{$\Delta$-settling}, in which the algorithm tries to settle visited vertices by scanning through their neighbors. The running time of the $\Delta$-settling random walk is then $\Oh( \Delta/\epsilon)$.

The algorithm handles edge insertions $\{u,v\}$ as follows:
if both endpoints are free, the edge is matched and the algorithm returns. 
The algorithm does nothing if both endpoints are matched.
If only one of the endpoints is matched, say $u$, then the algorithm unmatches $u$, sets $w:=\mate(u)$, and matches $\{u,v\}$.
Then a random walk as described above is started from $w$ to find an augmenting path (of length at most $\Oh(1/\epsilon)$).
If the random walk is unsuccessful, all changes are undone.
Deleting a matched edge $\{u,v\}$ leaves the two endpoints $u$~and~$v$ free. 
Then a random walk as described above is started from $u$ if it is free and from~$v$ if $v$ is free.
\begin{table*}[t!]
      \centering
      \small 
      \vspace*{-.5cm}
      \caption{Basic properties of the benchmark set of static graphs obtained from~\cite{benchmarksfornetworksanalysis,UFsparsematrixcollection,snap}.}
      \begin{tabular}{lrr@{\hskip 13pt} lrr@{\hskip 13pt} }
      graph & $n$ & $m$ & graph & $n$ & $m$ \\
              \midrule

144 &\numprint{144649} & \numprint{1074393} &               eu-2005 &\numprint{862664} & \numprint{16138468} \\
3elt &\numprint{4720} & \numprint{13722} &                  fe\_4elt2 &\numprint{11143} & \numprint{32818} \\
4elt &\numprint{15606} & \numprint{45878} &                 fe\_body &\numprint{45087} & \numprint{163734} \\ 
598a &\numprint{110971} & \numprint{741934} &               fe\_ocean &\numprint{143437} & \numprint{409593} \\ 
add20 &\numprint{2395} & \numprint{7462} &                  fe\_pwt &\numprint{36519} & \numprint{144794} \\
add32 &\numprint{4960} & \numprint{9462} &                  fe\_rotor &\numprint{99617} & \numprint{662431} \\ 
amazon-2008 &\numprint{735323} & \numprint{3523472} &       fe\_sphere &\numprint{16386} & \numprint{49152} \\ 
as-22july06 &\numprint{22963} & \numprint{48436} &          fe\_tooth &\numprint{78136} & \numprint{452591} \\ 
as-skitter &\numprint{554930} & \numprint{5797663} &        finan512 &\numprint{74752} & \numprint{261120} \\
auto &\numprint{448695} & \numprint{3314611} &              in-2004 &\numprint{1382908} & \numprint{13591473} \\
bcsstk29 &\numprint{13992} & \numprint{302748} &            loc-brightkite\_edges &\numprint{56739} & \numprint{212945} \\
bcsstk30 &\numprint{28924} & \numprint{1007284} &           loc-gowalla\_edges &\numprint{196591} & \numprint{950327} \\
bcsstk31 &\numprint{35588} & \numprint{572914} &            m14b &\numprint{214765} & \numprint{1679018} \\
bcsstk32 &\numprint{44609} & \numprint{985046} &            memplus &\numprint{17758} & \numprint{54196} \\
bcsstk33 &\numprint{8738} & \numprint{291583} &             p2p-Gnutella04 &\numprint{6405} & \numprint{29215} \\
brack2 &\numprint{62631} & \numprint{366559} &              PGPgiantcompo &\numprint{10680} & \numprint{24316} \\
citationCiteseer &\numprint{268495} & \numprint{1156647} &  rgg\_n\_2\_15\_s0 &\numprint{32768} & \numprint{160240} \\
cnr-2000 &\numprint{325557} & \numprint{2738969} &          soc-Slashdot0902 &\numprint{28550} & \numprint{379445} \\
coAuthorsCiteseer &\numprint{227320} & \numprint{814134} &  t60k &\numprint{60005} & \numprint{89440} \\
coAuthorsDBLP &\numprint{299067} & \numprint{977676} &      uk &\numprint{4824} & \numprint{6837} \\
coPapersCiteseer &\numprint{434102} & \numprint{16036720} & vibrobox &\numprint{12328} & \numprint{165250} \\
coPapersDBLP &\numprint{540486} & \numprint{15245729} &     wave &\numprint{156317} & \numprint{1059331} \\
crack &\numprint{10240} & \numprint{30380} &                web-Google &\numprint{356648} & \numprint{2093324} \\
cs4 &\numprint{22499} & \numprint{43858} &                  whitaker3 &\numprint{9800} & \numprint{28989} \\ 
cti &\numprint{16840} & \numprint{48232} &                  wiki-Talk &\numprint{232314} & \numprint{1458806}\\
data &\numprint{2851} & \numprint{15093} &                  wing &\numprint{62032} & \numprint{121544} \\
email-EuAll &\numprint{16805} & \numprint{60260} &          wing\_nodal &\numprint{10937} & \numprint{75488} \\
enron &\numprint{69244} & \numprint{254449} &               wordassociation-2011 &\numprint{10617} & \numprint{63788} \\
      \bottomrule
      \end{tabular}
      \label{staticgraphs}
\label{tab:graphstable}
\end{table*}

\begin{lemma}[Henzinger \etal \cite{DBLP:conf/esa/Henzinger0P020}]
\label{lemma:rw}
The random walk based algorithm maintains a $(1+\epsilon)$-approximate maximum matching if the length of the walk is at least $2/\epsilon-1$ and the walks are repeated $\Delta^{2/\epsilon-1} \log n$ times. 
\end{lemma}

\begin{corollary} If the algorithm by Stubbs and Williams uses the random walk-based MCM algorithm on each level s.t.~the length of the walks is $2/\epsilon-1$ and the walks are repeated $\Delta^{2/\epsilon-1} \log n$ times, the resulting \dynlevelR algorithm is $2(1+\epsilon)^2$-approximate with update time $\Oh(\epsilon^{-1}\Delta^{2/\epsilon-1} \log n (\log_{1+\epsilon}N/L)^2)$.
\end{corollary}

\paragraph*{Blossom-based MCM.}
Roughly speaking, the algorithm runs a modified BFS that is bounded in depth by $2/\epsilon -1$ to find augmenting paths from a free node. This has a theoretically faster running time of $\Oh(\Delta^{2/\epsilon-1})$ per free node and still guarantees a $(1+\epsilon)$ approximation.
The algorithm handles insertions as follows. Let $\{u,v\}$ be the inserted edge. If $u$ and $v$ are free, then the edge is matched directly and the algorithm stops. Otherwise, an augmenting path search is started from $u$ if $u$ is free and from $v$ if $v$ is free.  
If both $u$ and $v$ are not free, then a breadth-first search from $u$ is started to find a free node reachable via an alternating path. From this node an augmenting path search is started. 
If the algorithm does not handle this case, \ie if both $u$ and $v$ are not free, it does nothing and is called \emph{unsafe}. If the algorithm handles the this case and the augmenting path search is not depth-bounded, then it maintains a maximum matching.
The algorithm handles deletions as follows: let $\{u,v\}$ be the deleted edge. After the deletion, an augmenting path search is started from any free endpoint $u$ or $v$. 
The algorithm uses two optimizations: \emph{lazy augmenting path search}, in which an augmenting path search from $u$ and $v$ is only  started if at least $x$ edges have been inserted or deleted since the last augmenting path search from $u$ or $v$ or if no augmenting path search has been started so far. 
The second optimization limits the search depth of the augmenting path search to $2/\epsilon-1$. This ensures that there is no augmenting path of length at most $2/\epsilon -1$, thus yielding a deterministic $(1+\epsilon)$-approximate matching algorithm (if the algorithm is run with the \emph{safe} option). 

\begin{corollary} If the algorithm by Stubbs and Williams uses the blossom-based MCM algorithm on each level s.\,t.\ the depth is bounded by $2/\epsilon-1$, then the resulting \dynlevelB algorithm is $2(1+\epsilon)^2$-approximate with update time $\Oh(\Delta^{2/\epsilon-1}(\log_{1+\epsilon}N/L)^2)$.
\end{corollary}

\section{Experiments}
\paragraph*{Implementation and Compute System.}\label{Methodology}
We implemented the algorithms described above in \texttt {C++}
compiled with \texttt{g++-9.3.0} using the optimization flag \texttt{-O3}. 
For the dynamic MCM algorithms, we use the implementation by Henzinger \etal~\cite{DBLP:conf/esa/Henzinger0P020}.
All codes are sequential.
We intend to release our implementation after paper acceptance.

Our algorithms use the following dynamic graph data structure. For each node $u$, we maintain a vector $L_u$ of adjacent nodes, and a hash table $\mathcal{H}_u$ that maps a neighbor $v$ of $u$ to its position in $L_u$. 
This data structure allows for expected constant time insertion and deletion as well as a constant time operation to select a random neighbor of $u$. 
The deletion operation on $\{u,v\}$ is implemented as follows: get the position of $v$ in $L_u$ via a lookup in $\mathcal{H}_u(v)$. 
Swap the element in $L_u$ with the last element $w$ in the vector and update the position of $w$ in $\mathcal{H}_u$. 
Finally, pop the last element (now $v$) from $L_u$ and delete its entry from $\mathcal{H}_u$.

Our experiments are conducted on one core of a
machine with a AMD EPYC 7702P 64-core CPU (2~GHz) and 1~TB of RAM.
In order to be able to compare our results with optimum weight matchings, we use the LEMON library~\cite{DEZSO201123}, which provides algorithms for static weighted maximum matching.

\paragraph*{Methodology.}
By default we perform ten repetitions per instance.
We measure the total time taken to compute all edge insertions and deletions
and generally use the \emph{geometric mean} when averaging over different instances
in order to give every instance a comparable influence on the final result. 
In order to compare different algorithms, we use \emph{performance profiles}~\cite{DBLP:journals/mp/DolanM02}.
These plots relate the matching weight of all algorithms to the optimum matching weight.
More precisely, the $y$-axis shows $\#\{\textit{objective} \geq \tau \times \textit{OPT} \} / \# \text{instances}$, where \textit{objective} corresponds to
the result of an algorithm on an instance and \textit{OPT} refers to the optimum result.
The parameter $\tau\leq 1$ in this equation is plotted on the $x$-axis.
For each algorithm, this yields a non-decreasing, piecewise constant function.
\begin{figure*}[t]
\vspace*{-1cm}
\centering
\includegraphics[width=.44\textwidth]{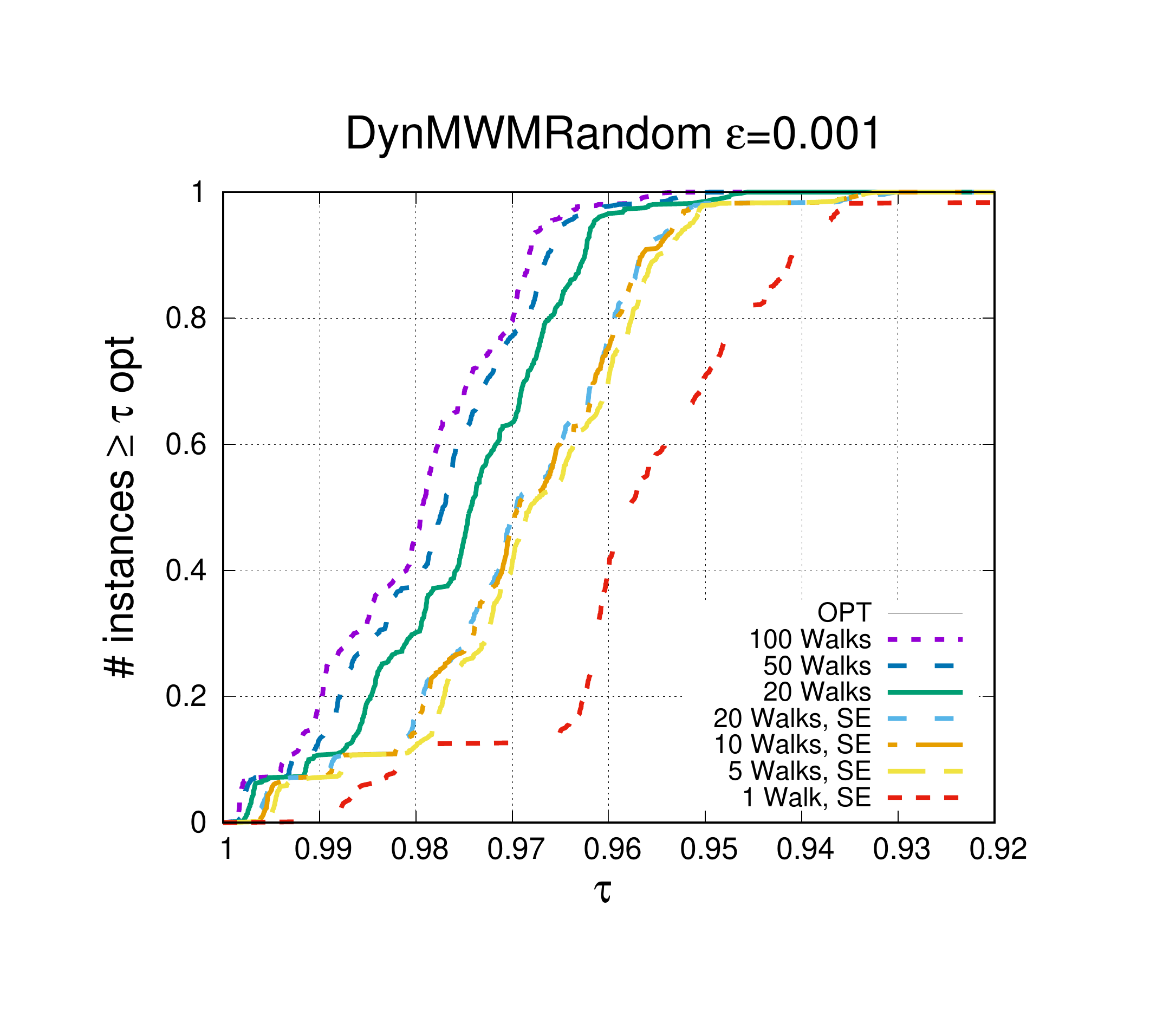} 
\includegraphics[width=.44\textwidth]{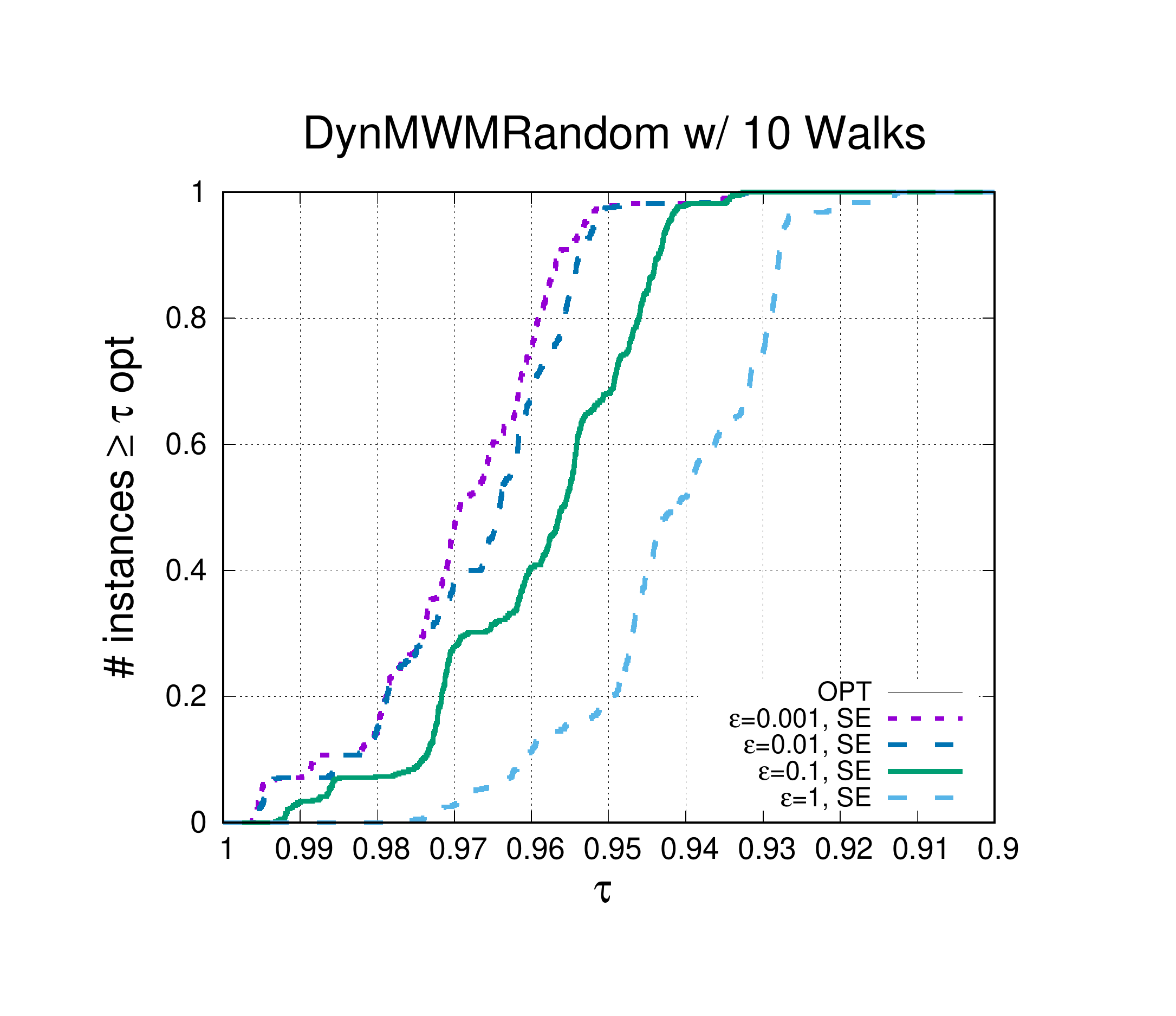}
\vspace*{-1cm}
\caption{Left: Number of random walks and random walks with ``stop early'' heuristic (SE) for fixed epsilon. Right: Various values of $\epsilon$ for fixed number of walks.}
\vspace*{-.5cm}
\label{fig:rwrepetitions}
\end{figure*}

\begin{table}[tb]
      \centering
      \small 
\vspace*{-.25cm}
      \caption{Basic properties of the benchmark set of dynamic graphs with number of update operations $\mathcal{U}$. Most of the graphs only feature insertions. The only two exceptions are marked with a *. All of these graphs have been obtained from the KONECT graph database~\cite{konect:unlink}.} 
      \begin{tabular}{lrr@{\hskip 13pt}}
      graph & $n$ & $\mathcal{U}$ \\
\midrule
\midrule
amazon-ratings &\numprint{2146058} & \numprint{5838041} \\ 
citeulike\_ui &\numprint{731770} & \numprint{2411819} \\ 
dewiki$^*$ &\numprint{2166670} & \numprint{86337879} \\ 
dnc-temporalGraph &\numprint{2030} & \numprint{39264} \\ 
facebook-wosn-wall &\numprint{46953} & \numprint{876993} \\ 
flickr-growth &\numprint{2302926} & \numprint{33140017} \\ 
haggle &\numprint{275} & \numprint{28244} \\ 
lastfm\_band &\numprint{174078} & \numprint{19150868} \\ 
lkml-reply &\numprint{63400} & \numprint{1096440} \\ 
movielens10m &\numprint{69879} & \numprint{10000054} \\ 
munmun\_digg &\numprint{30399} & \numprint{87627} \\ 
proper\_loans &\numprint{89270} & \numprint{3394979} \\ 
sociopatterns-infections &\numprint{411} & \numprint{17298} \\ 
stackexchange-overflow  &\numprint{545197} & \numprint{1301942} \\ 
topology &\numprint{34762} & \numprint{171403} \\ 
wikipedia-growth &\numprint{1870710} & \numprint{39953145} \\ 
wiki\_simple\_en$^*$ &\numprint{100313} & \numprint{1627472} \\ 
youtube-u-growth &\numprint{3223590} & \numprint{9375374} \\
\bottomrule
\end{tabular}
\vspace*{-.5cm}
\label{dyninstances}
\end{table}

\paragraph*{Instances.} We evaluate our algorithms on graphs collected 
from various resources~\cite{benchmarksfornetworksanalysis,UFsparsematrixcollection,snap,DBLP:conf/www/Kunegis13,konect:unlink,DBLP:journals/corr/abs-2102-11169}, which are also used in related studies~\cite{DBLP:conf/esa/Henzinger0P020}.
      Table~\ref{tab:graphstable} summarizes the main properties of the benchmark set.
      Our benchmark set includes a number of graphs from numeric simulations as well as complex networks.
These include static graphs as well as real dynamic graphs.
As our algorithms only handle undirected simple graphs, we ignore possible edge directions in the input graphs, and we remove self-loops and parallel edges.
We perform \emph{two different types} of experiments. 
First, we use the algorithms using insertions only, \ie we start with an empty graph and insert all edges of the static graph in a random order. We do this with all graphs from~Table~\ref{staticgraphs}.
Second, we use real dynamic instances from Table~\ref{dyninstances}. Most of these instances, however, only feature insertions (exceptions: \texttt{dewiki} and \texttt{wiki-simple-en}). 
Hence, we perform additional experiments with fully-dynamic graphs from these inputs, by undoing $x$ percent of the update operations performed last.
As the instances are unweighted, we generated weights for edges uniformly at random in $[1, 100]$.

\subsection{\dynrand.}
First, we look at \dynrand (in various configurations) on insertions-only instances. We start with the effect of the number of walks done after each update and the impact of the ``stop early'' heuristic. 
In Fig~\ref{fig:rwrepetitions} (left), we look at different numbers of walks when (not) using the ``stop early''
heuristic (if used, $\epsilon$ is set to $10^{-3}$). Geometric mean values for running time and achieved matching weight are in Table~\ref{tab:geomeanfixedeps}.
As expected, increasing the number of walks increases the weight of the computed matchings (with and without ``stop early'' heuristic). 
With the ``stop early'' heuristic (in our experiments $\beta = 5$), increasing the number of walks more than ten does not significantly improve the solutions. 
When disabling the ``stop early'' heuristic, solutions improve again, even for the same number of 20 walks.  
The strongest configuration in this setting uses 100 walks with the ``stop early'' heuristic being enabled.  The configuration is as close as 4\% to the optimum  on more than 95\% of the instances. 
Note that the number of walks done here is much less than what Lemma~\ref{lemma:rw} suggests to achieve a $(1+\epsilon)$-approximation.
Moreover, using more walks as well as not using the ``stop early'' heuristic also comes at the cost of additional running time. 
For example, for $\epsilon=0.001$ and $20$ walks, using the ``stop early'' heuristic saves roughly a factor of three in running time.  
We conclude that for a fixed $\epsilon$, the number of walks and the ``stop early'' heuristic yield a favorable speed quality trade-off.

\begin{table}[tb]
\vspace*{-.25cm}
\caption{Top: Geometric mean running times and weight of the final matching for fixed $\epsilon=0.001$ and varying number of walks. Bottom: Geometric mean running times and weight of final matching for fixed number of walks (10) and varying $\epsilon$.}
\label{tab:geomeanfixedeps}
\label{tab:geomeanfixedreps}
\vspace*{.25cm}
\centering
\small 
\begin{tabular}{r|r|r}
                \toprule
\# walks & geo. mean $t$ [s] & geo. mean $\omega(\mathcal{M})$ \\
                \midrule
\multicolumn{3}{c}{with SEHeuristic} \\
                \midrule
1 &4.0  & 1572285.92 \\ 
5 &16.4  & +1.15\%\\ 
10& 20.5  & +1.28\%\\
20& 21.8  & +1.31\%\\
                \midrule
\multicolumn{3}{c}{without SEHeuristic} \\
                \midrule
20 &56.4 &+1.91\%\\
50 &128.6  &+2.24\%\\
100 &241.6 &+2.44\%  \\
                \midrule
\midrule
$\epsilon$ & geo. mean $t$ [s] & geo. mean $\omega(\mathcal{M})$ \\
                \midrule
\multicolumn{3}{c}{with SEHeuristic} \\
                \midrule
1         & 2.4  & 1547129.56 \\
$10^{-1}$ & 5.9 & +1.84\%\\
$10^{-2}$ & 13.7 & +2.71\%\\
$10^{-3}$ & 20.5 & +2.93\%\\

      \bottomrule
\end{tabular}
\vspace*{-.35cm}
\end{table}

Next, we fix the use of the ``stop early'' heuristic and the number of walks per update to ten. Figure~\ref{fig:rwrepetitions} (right) shows the performance compared to the optimum solution value and Table~\ref{tab:geomeanfixedreps} gives geometric mean values for running time and matching weight again.
When using $\epsilon=1$, the algorithm computes in all cases at least a matching that has $91.2$\% of the optimum weighted matching weight. As soon as we decrease $\epsilon$, solutions improve noticeably. 
For the case $\epsilon=10^{-1}$, solutions improve by $1.84$\% on average. 
The algorithm always computes a matching which has at least 93.3\% of the optimum weight matching. However, the parameter $\epsilon$ seems to be exhaustive in the sense that decreasing it from $10^{-2}$ to $10^{-3}$ does not significantly improve solutions anymore. 
 Also note that the running time grows much slowlier than $1/\epsilon$.
This is due to the ``stop early'' heuristic. Overall, we conclude here that both the number of walks and the parameter $\epsilon$ are very important to maintain heavy matchings.

On the other hand, it is important to mention that in the unweighted case the quality of the algorithms based on random walks 
is not very sensitive to both parameters, $\epsilon$ and the number of walks~\cite{DBLP:conf/esa/Henzinger0P020}.
That is why Henzinger~\etal~\cite{DBLP:conf/esa/Henzinger0P020} use only one random walk and vary only~$\epsilon$ in their final MCM algorithm.
The different behavior comes from two observations: first, in the unweighted case many subpaths of an augmenting path are often interchangeable. In the weighted case, this may not be the case and hence it can be hard to find some of the remaining augmenting paths in later stages of the algorithms. Moreover, the algorithms differ in the way the random walkers work. 
Our weighted algorithm builds a path and then solves a dynamic program on it (hence, the length of a random walker is limited by $n$). In the unweighted case, a random walker does changes to the graph on the fly and undoes changes in the end if it was unsuccessful. Hence, it can in principle run much longer than $n$ steps.

In terms of running time, we can give a rough estimate of a speedup that would be obtained when running the optimal algorithm after each update. Note that the comparison is somewhat unfair since we compare an optimal algorithm with a heuristic algorithm and since we did not run the optimum algorithm after each update step due to excessive time necessary to run this experiment. On the largest instance of the set, eu-2005, the time per update of the slowest dynamic algorithm in these experiments ($\epsilon=0.001$, 100 walks, no stop heuristic) is 0.00227s on average. On the other hand, the optimum algorithm needs 79.8998s to compute the solution on the final graph. This is more than four orders of maginitude difference. On the other hand, the fastest algorithms in this experiment are at least one additional order of magnitude faster.
\label{exp:rw}

\subsection{\dynlevel.}  
We now evaluate the performance of our implementation \dynlevel of the meta-algorithm by Crouch, Stubbs, and Williams.
First, we use $\alpha=(1+\epsilon)$, so the algorithms under investigation have approximation guarantee of $2(1+\epsilon)^2$. Note that decreasing $\epsilon$ increases both, the number of levels in the algorithm necessary as well as the work performed by the unweighted matching algorithms on each level of the hierarchy.
The results for the meta-algorithm are roughly the same when using the unweighted random walk algorithm compared to using the unweighted blossom-based algorithm for the same value of $\epsilon$ on each level.
For example, the difference between \dynlevelR and \dynlevelB is smaller than 0.5\% on average for $\epsilon=0.1$ (with random walks having the slight advantage). Moreover, both instantiations have roughly the same running time on average (less than 1.6\% difference on average).  Hence, we focus our presentation on \dynlevelR.
\begin{figure}[t]
\vspace*{-1cm}
\centering
\includegraphics[width=.44\textwidth]{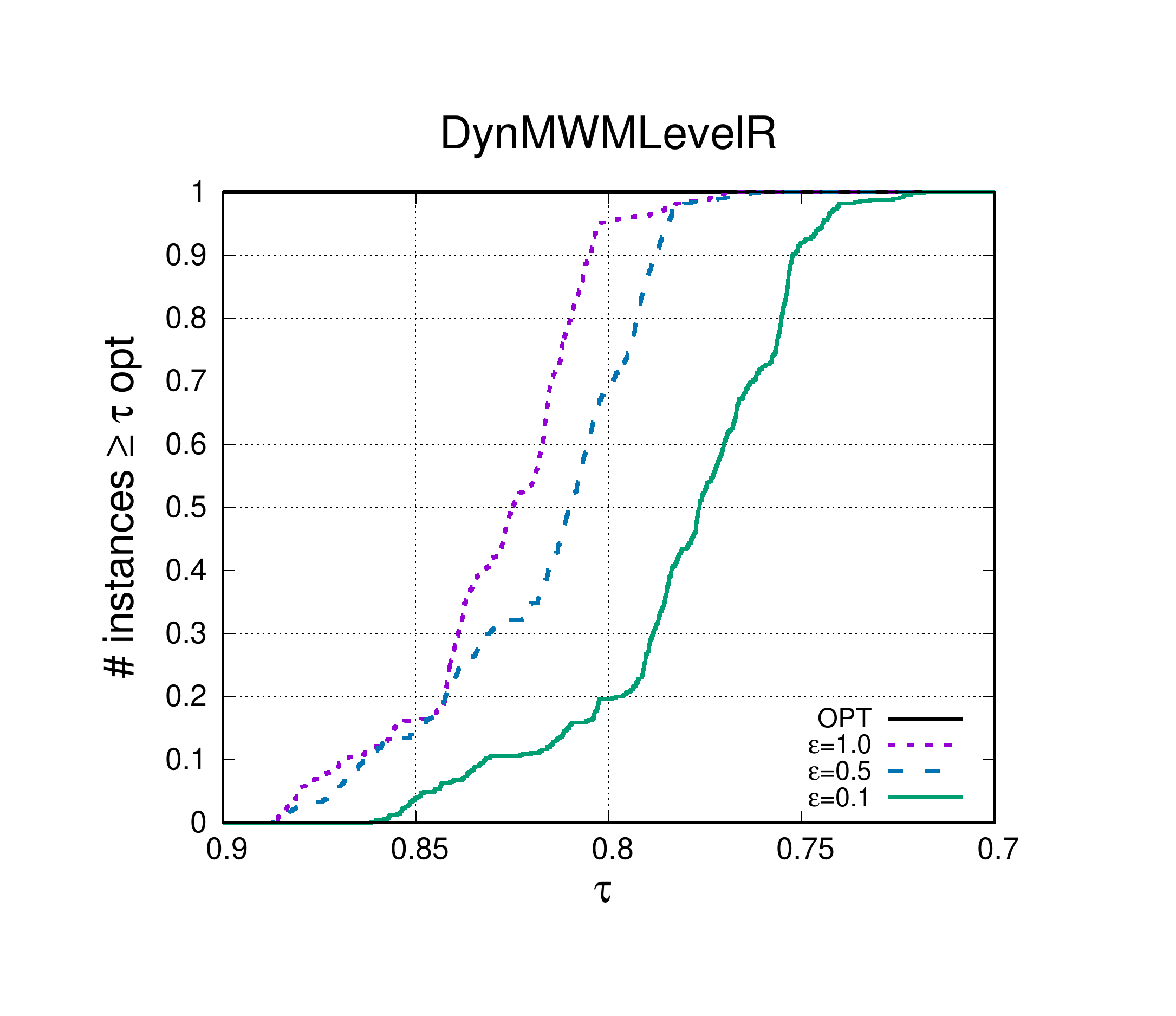}
\vspace*{-1cm}
\caption{Performance profile of the meta-algorithm \dynlevelR instantiated with random-walk based algorithms as MCM algorithms for different values of $\epsilon$.}
\label{fig:crubbs}
\vspace*{-.5cm}
\end{figure}
Figure~\ref{fig:crubbs} shows three different configurations of the algorithm using different values of $\epsilon \in \{1,0.5,0.1\}$. We do not run smaller values of $\epsilon$ as the algorithm creates a lot of levels for small $\epsilon$ and thus needs a large amount of memory. This is due to the large amount of levels created for small values of $\epsilon$. To give a rough estimate, on the smallest graph from the collection, add20, \dynlevelR with $\epsilon=0.1$ needs more than a factor 40 more memory than \dynrand.

First, note that the results are significantly worse than the results achieved by \dynrand: \dynlevelR computes at least a weighted matching that has $71.8$\% of the optimum weight matching for $\epsilon=0.1$. Note that this is very well within the factor 2 that theory suggests. 
\begin{figure*}[t!]
\centering
\includegraphics[width=.44\textwidth]{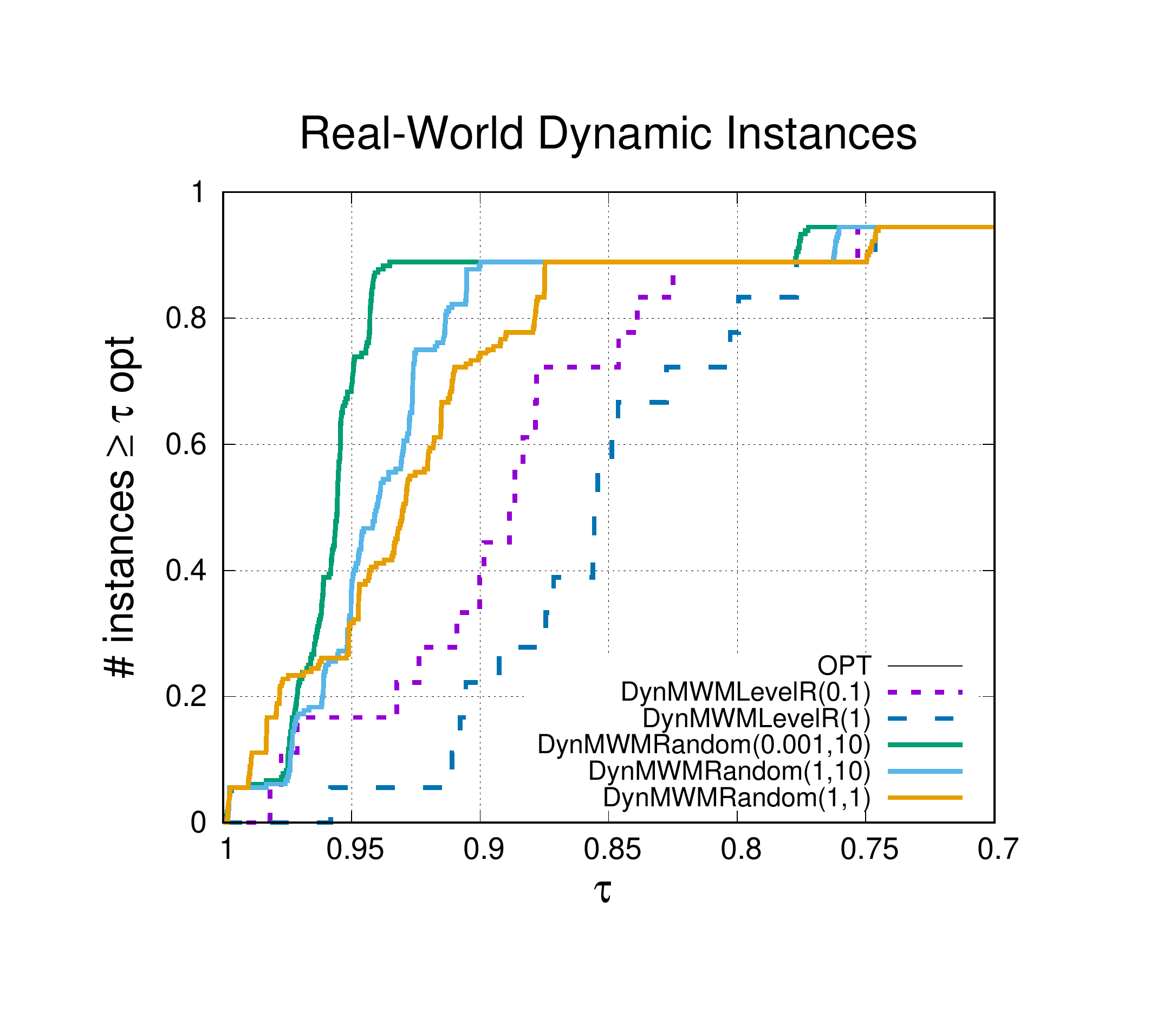}
\includegraphics[width=.44\textwidth]{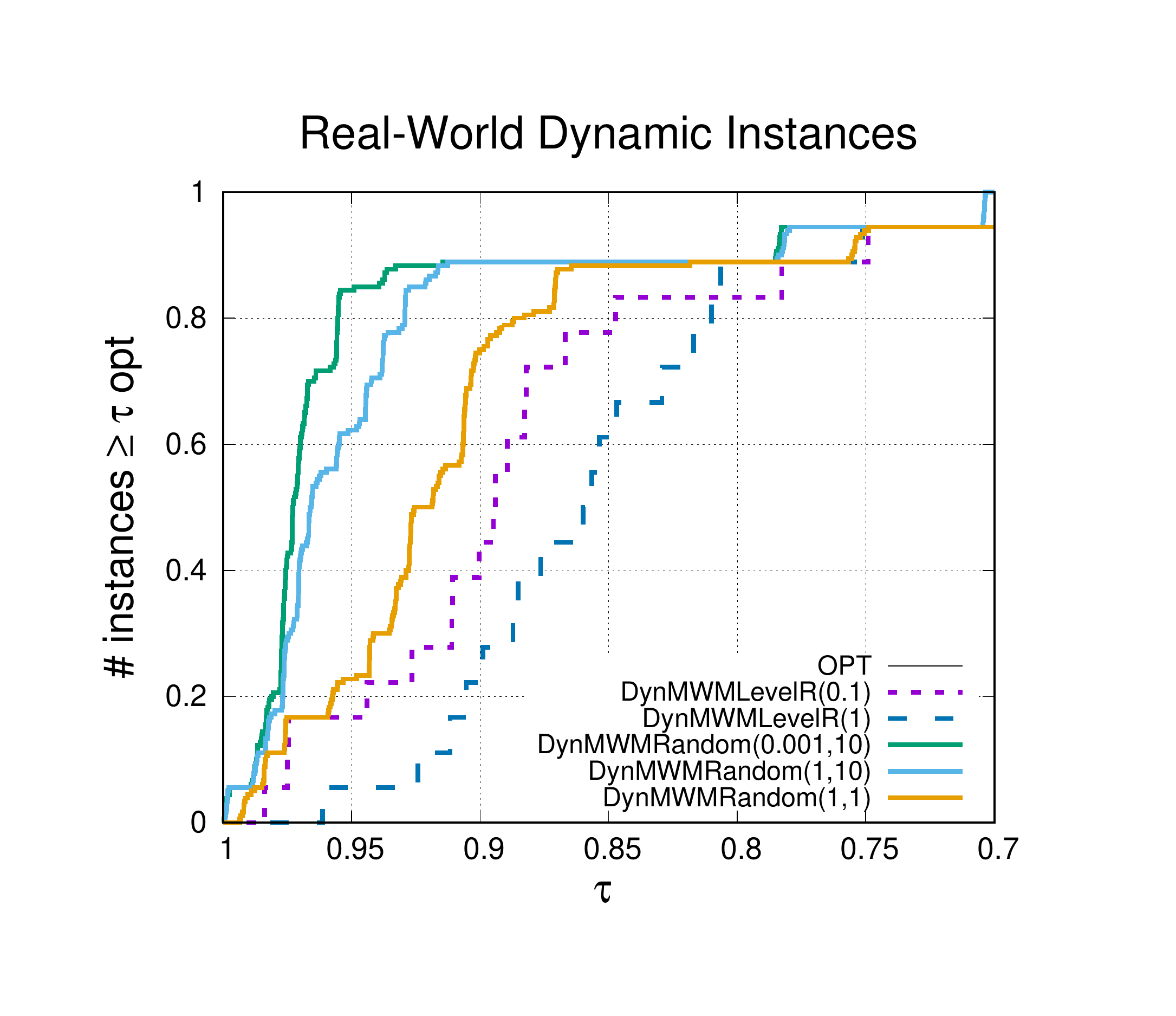}
\vspace*{-1cm}
\caption{Comparison of different algorithm configuration on real-world dynamic instances without (left) and with (right) 25\% undo operations.}
\vspace*{-.5cm}
\label{fig:rwdinst}
\end{figure*}
However, for larger values of $\epsilon$, the algorithm computes even better results. For $\epsilon=0.5$ the algorithm computes a weighted matching of at least $76.1\%$ of the optimum weight matching, and for $\epsilon=1$, the algorithm computes $76.8$\% of the optimum weight matching.  We believe that this somewhat surprising behavior is due to two reasons.  
First, if one only increases the work done by the MCM algorithms on each level, then the cardinality of the matchings on each level increases as expected. However, the weight of the computed matching on each level does not necessarily increase. Hence, the level-based meta-algorithm has a too strong preference on the cardinality as it ignores the weights on each level. Moreover, when increasing the number of levels, the priority of heavy edges increases, as there are only very few of them at the highest levels in the hierarchy. Hence, the algorithm becomes closer to a greedy algorithm that picks heavy edges first.

In terms of running time, \dynlevelR uses 1.53s, 2.66s, and 14.61s on average for $\epsilon=1$, $0.5$, and $0.1$, respectively. Moreover, the best and fastest configuration in this category uses $\epsilon=1$, computes 11\% smaller matchings than the augmented random walk for $\epsilon=1$ (recall that the number of steps of the walk is $2/\epsilon+3)$) and the geometric mean running time of the random walk is 0.61s. We conclude here that \dynrand is better and faster than \dynlevel. Moreover, augmented random walks need significantly less memory.

\subsection{Real-World Dynamic Instances.}
We now switch to the real-world dynamic instances. Note that most of these instances are insertion-only (with two exceptions); however, the insertions are ordered in the way they appear in the real world.
Hence, we perform additional experiments with fully-dynamic graphs from these inputs, by undoing $x$ percent of the update operations performed last (call them  $\mathcal{U}_x$). 
More precisely, we perform the operations in $\mathcal{U}_x$ in reverse order. If an edge operation was an insertion in $\mathcal{U}_x$, we perform a delete operation and if it was a delete operation, we insert it.
However, when reinserting an edge, we use a new random weight. In our experiments, we perform 0\%, 10\% and 25\%  undo operations.  As before, we compute the update on the graph after each removal/insertion. 
Table~\ref{tab:geomeanrealdyn} summarizes the results of the experiment and Figure~\ref{fig:rwdinst} compares the final quality of the dynamic algorithms with the optimum on the final graph after all updates have been performed.
\begin{table}[tb]
\vspace*{-.25cm}
\caption{Geometric mean results for real-world dynamic instances with and without undo operations depending on $\epsilon$ and walks per update ($w$). Smaller is better. }
\vspace*{.25cm}
\label{tab:geomeanrealdyn}
\centering
\small 
\begin{tabular}{l|r|r}
                \toprule
Algorithm  & $\overline{t[s]}$ & $\overline{OPT/\omega(\mathcal{M})}$ \\
                \midrule
\multicolumn{3}{c}{without undo ops} \\
                \midrule
Random($\epsilon=10^{-3}$, $w$=10)  & 11.2 & 7.3\%\\        
Random($\epsilon=1$, $w$=10)          & 3.9  & 9.0\%\\        
Random($\epsilon=1$, $w$=1)           & 1.0  & 10.0\%\\        
LevelR($\epsilon=10^{-1}$)    & 17.3 & 14.0\%\\        
LevelR($\epsilon=1$)            & 1.7  & 18.7\%\\        
\midrule
\multicolumn{3}{c}{with 10\% undo ops} \\
\midrule
Random($\epsilon=10^{-3}$, $w$=10) & 14.3  & 6.2\%\\      
Random($\epsilon=1$, $w$=10)         & 5.0    &7.2\% \\     
Random($\epsilon=1$, $w$=1)         & 1.2    &10.7\% \\     
LevelR($\epsilon=10^{-1}$)   & 18.6   &13.5\% \\     
LevelR($\epsilon=1$)           & 1.9    &17.9\% \\     
\midrule
\multicolumn{3}{c}{with 25\% undo ops} \\
\midrule
Random($\epsilon=10^{-3}$, $w$=10) & 18.1   & 6.0\% \\     
Random($\epsilon=1$, $w$=10)         & 6.3    & 7.0\%\\     
Random($\epsilon=1$, $w$=1)          & 1.5    & 10.9\%\\     
LevelR($\epsilon=10^{-1}$)   & 20.5    &13.8\% \\     
LevelR($\epsilon=1$)           & 2.1   & 17.5\%\\     

      \bottomrule
\end{tabular}
\vspace*{-.5cm}
\end{table}
For our experiments with \dynrand, we use the configurations $\epsilon=10^{-3}$ with 10 walks, $\epsilon=1$ with 10 walks, and $\epsilon=1$ with 1 walk. In any case, for these algorithms the ``stop early'' heuristic is enabled. 
For the \dynlevelR algorithm, we use $\epsilon=10^{-1}$ and $\epsilon=1$. 
Overall, the results confirm the results of the previous section. For example, \dynrand($\epsilon=0.001$, 10 walks per update), computes matchings with 93.5\% weight of the optimum weight matching on more than 89\% of the instances. The fastest configuration \dynrand($\epsilon=1$, 1 walk per update), is also already very good. It computes 90\% of the optimum weight matching on 89\% of the instances. On the other hand, \dynlevelR computes 82.5\% and 77.7\% of the optimum weight matching on 88.9\% of the instances for $\epsilon=0.1$ and $\epsilon=1$, respectively. 

Moreover, the experiments indicate that \dynlevelR is always outperformed in terms of quality and speed by a configuration of \dynrand. 
This is true with and without running undo operations. For example, without running undo operations, \dynlevelR($\epsilon=10^{-1}$) computes solutions that are 14\% smaller than the optimum matching weight on average while taking 17.3s running time. 
On the other hand, already \dynrand($\epsilon=1$, 1 walk per update) computes weighted matchings that are 10\% worse than the optimum solution while needing only 1s running time on average. The same comparison can be done when undo operations are performed, \ie \dynlevelR($\epsilon=10^{-1}$) is always outperformed by \dynrand($\epsilon=1$, 1 walk). 

When considering undo operations, the quality of all algorithms is stable, \ie there is no reduction in quality observable over the case without undo operations.
Note that in contrast the quality in terms of distance to OPT often improves when undo operations are performed. This is due to the fact that the algorithms perform additional work when the undo operations are called. 
Hence, the results may find additional augmenting paths that had not been found when the algorithm was at the original state of the graph.

\section{Conclusions and Future Work}
\label{sec:conclusions}
Our experiments show that fully-dynamic approximation algorithms for the MWM problem are viable alternatives
compared to an exact approach. Of the algorithms we implemented, \dynrand fares best overall and should be the
method of choice whenever exact solutions are unnecessary. After all, the solutions are often close to the
optimum.
Future work could focus on finding augmenting paths of length $3$ efficiently, \eg\ as suggested in~\cite{NeimanS16}
for the unweighted case.

\FloatBarrier
\pagebreak

\bibliographystyle{plainurl}
\bibliography{phdthesiscs}
\vfill 
\clearpage
\newpage
\appendix

\section{Pseudocode of Dynamic Programming for MWM on Paths}
\label{app:sec:dynprogalgo}
\begin{algorithm}[h!]
\begin{algorithmic}[1]
        \STATE $W[0] := 0; W[1] := \omega(e_1)$
        \STATE $M[0] := \emptyset; M[1] := \{e_1\}$
        \STATE \textbf{for} $i := 2$ \textbf{to} $k$ \textbf{do}
        \STATE \quad \quad \textbf{if} $\omega(e_i) + W[i-2] > W[i-1]$ \textbf{then}
        \STATE \quad \quad \quad \quad $W[i] := \omega(e_i) + W[i-2]$
        \STATE \quad \quad \quad \quad $M[i] := M[i-2] \cup \{e_i\}$
        \STATE \quad \quad \textbf{else} 
        \STATE \quad \quad \quad \quad $W[i] := W[i-1]$
        \STATE \quad \quad \quad \quad $M[i] := M[i-1]$       
        \RETURN $M[k]$
\end{algorithmic}
\caption{MWM($\mathcal{P}=\langle e_1, \ldots, e_k \rangle$)~\cite{MauSan07}}
\label{algo:dynprog}
\end{algorithm}

\section{Omitted Proofs}
\label{appendixproof}

\subsection{Proof of Proposition~\ref{thm:kappx}}
~

\noindent A {\textit{weight-augmenting path}} $\mathcal{P}$ with respect to a matching $\mathcal{M}$ is an alternating path with respect to $\mathcal{M}$,
and the matching $\mathcal{M}\oplus \mathcal{P}$ has a larger weight than $\mathcal{M}$; 
in other words $w(\mathcal{M}\oplus \mathcal{P})> w(\mathcal{M})$.
A weight-augmenting path with $k$ edges outside $\mathcal{M}$ is called {\textit{weight-augmenting $k$-path}}.
A weight-augmenting $k$-path thus can have $2k-1, 2k,$ or $2k+1$ edges, 
whereas in the unweighted case an augmenting path with $k$ edges outside of a given matching should have exactly $2k-1$ edges. 
The following proposition writes the approximation guarantee of a matching in terms of weight-augmenting $k$-paths.
\begin{figure*}
\centerline{\includegraphics[width=0.8\textwidth]{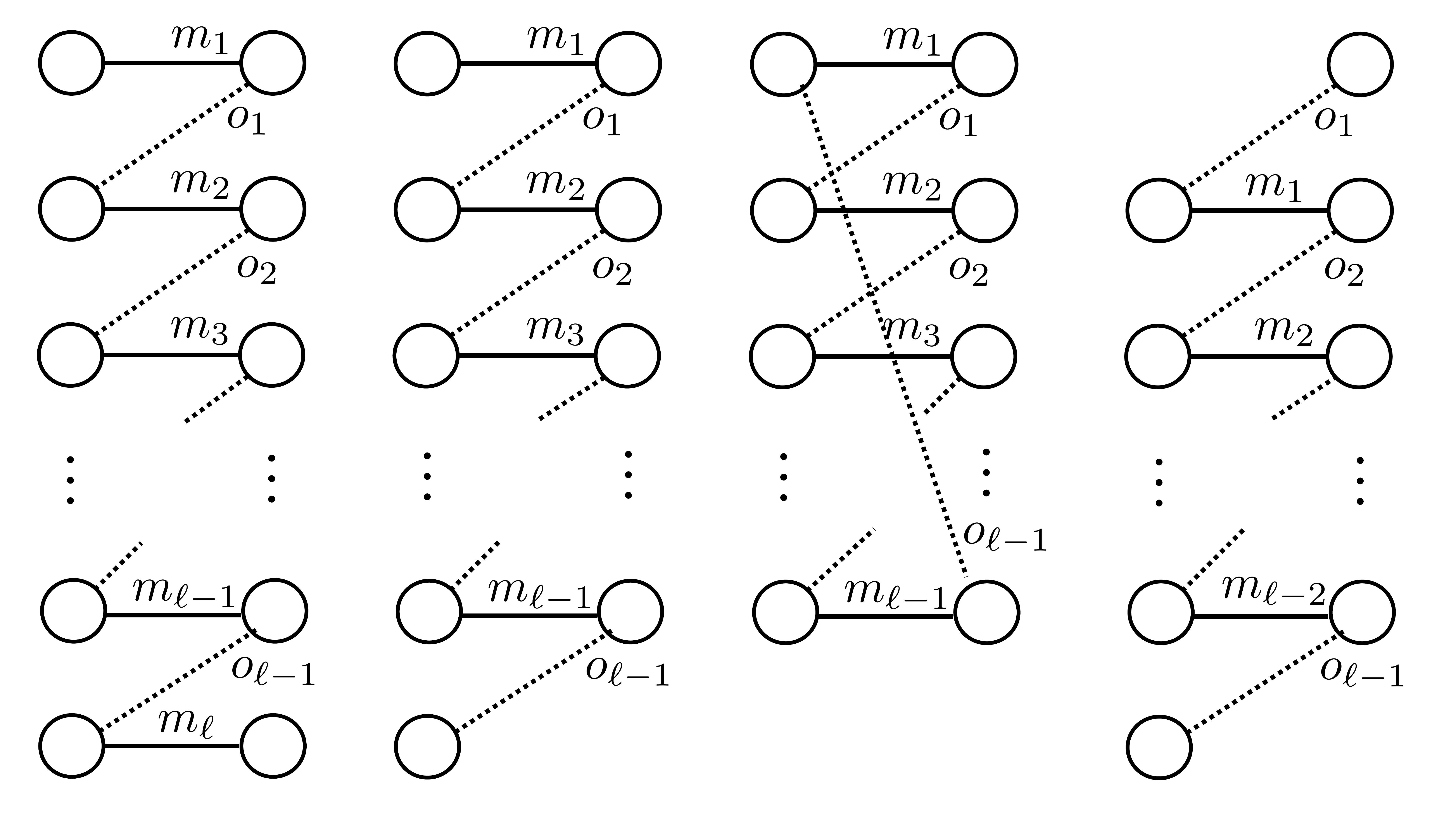}}
\caption{Four different type of components of $\mathcal{M}\oplus\mathcal{M}^*$.
The edges labeled with $m_i$ belong to $\mathcal{M}$ and are shown with the solid lines.
The edges labeled with $o_j$ belong to $\mathcal{M}^*$ and are shown with the dotted lines.
There are other edges of  $\mathcal{M}$ and $\mathcal{M}^*$ which are not shown.
If $k$ is the smallest number for which is a weight-augmenting $k$-path, then $\ell>k$.
\label{fig:mopluso}}
\end{figure*}

\begin{proposition}
\label{th:kappx}
Let $\mathcal{M}$ be a matching and $k\geq 1$ be the smallest number
such that there is a weight-augmenting $k$-path with respect to $\mathcal{M}$.
Then $w(\mathcal{M})\geq \frac{k-1}{k} \cdot w(\mathcal{M}^*)$, where $\mathcal{M}^*$ is a matching with the maximum weight.
\end{proposition}

Before proving the proposition, we note that its result is known.
Drake and Hougardy~\cite{drho:03} show a version of the proposition
for weight-augmenting $3$-paths.
Pothen et al.~\cite[Lemma 3.1]{pofm:19} generalize the proof given by Drake and Hougardy 
and show the result given in \Cref{th:kappx}.
Our proof is somewhat different from the previous one, 
is closer to the way we make us of the result in the random walk based algorithm,
and is therefore given for completeness.

\begin{proof}[\Cref{th:kappx}]
Consider the symmetric difference  $\mathcal{M}\oplus \mathcal{M}^*$, which 
can have four different type of components displayed in Figure~\ref{fig:mopluso}.
In the figure, the edges labeled with $m_i$ belong to $\mathcal{M}$ and are shown with the solid lines.
The edges labeled with $o_j$ belong to $\mathcal{M}^*$ and are shown with the dotted lines.
There are other edges of  $\mathcal{M}$ and $\mathcal{M}^*$ which are not shown.
In the leftmost component, there is one more edge in $\mathcal{M}$ than in $\mathcal{M}^*$; in the second and third one, there are 
equal number of edges in $\mathcal{M}$ and $\mathcal{M}^*$; in the rightmost one, there is one more edge in $\mathcal{M}^*$ than in $\mathcal{M}$.
In any of these components, $\ell > k$, if there is to be any weight-augmenting paths.

We now show that in a component $C$ with at least $k$ edges from $O$, we have $w(\mathcal{M}\cap C)\geq \frac{k-1}{k}w(\mathcal{M}^*\cap C)$.
This will show that in all components of $\mathcal{M}\oplus \mathcal{M}^*$, we have the same guarantee of
$\frac{k-1}{k}$ and hence the result.

Let us start with the leftmost component of Figure~\ref{fig:mopluso}.
We will use $m_i$ and $o_j$ for the weight of the corresponding edges.
We have $$m_1+m_2\geq o_1$$ otherwise, there will be a weight-augmenting 1-path.
We can continue with $$m_1+m_2+m_3\geq o_1+o_2$$ with the same reasoning 
until
$$m_1+m_2+\cdots +m_k\geq o_1+ o_2+\cdots+ o_{k-1}\;,$$
where we have $k$ edges of $\mathcal{M}$ on the left-hand side of the equation 
and $k-1$ edges of $\mathcal{M}^*$ on the right-hand side of the equation.
At this point, we have included $m_1$ and $m_2$ in $k-1$ inequalities,
$m_3$ in $k-2$ inequalities, and in general $m_j$ for $3\leq j\leq k$ in $k-j+1$ inequalities.
Then we keep the same number of edges for another $\ell - k$ steps and obtain
the inequalities
\begin{align*}
m_2+m_3+\cdots+ m_{k+1}&\geq o_2+o_3+\cdots+ o_{k}\;,\\
m_3+m_4+\cdots +m_{k+2}&\geq o_3+ o_4+\cdots+o_{k+1}\;,\\
&\vdots\\
m_{\ell-k+1}+\cdots+m_{\ell}&\geq  o_{\ell-k+1}+\cdots+o_{\ell-1}\;,
\end{align*}
as none of these $(k-1)$-paths are weight-augmenting.
In these inequalities, $m_j$ for $2\leq j\leq \ell-k+1$ appears as the first term once,
and it is included in the equalities starting with $m_i$ for all $j-k+1\leq i<j$.
Up to this point, $m_j$ for $2\leq j \leq \ell-k+1$ is included in $k$ inequalities.
Once we have included the last edge $m_{\ell}$ of $\mathcal{M}$ in an inequality,
we continue with writing down new inequalities by removing one edge at a time from the left and the right of the last inequality, 
starting from the 
edge with the smallest index:
\begin{align*}
m_{\ell-k+2}+\cdots+m_{\ell}&\geq  o_{\ell-k+2}+\cdots+o_{\ell-1}\\
m_{\ell-k+3}+\cdots+m_{\ell}&\geq  o_{\ell -k+3}+\cdots+o_{\ell-1}\\
&\vdots\\
m_{\ell-1}+m_{\ell}&\geq  o_{\ell-1}\;,
\end{align*}
for otherwise there would be a weight-augmenting $j$-path with a $j < k$.
We see that the last $k$ edges of $\mathcal{M}$ are symmetrical 
to the first $k$ edges of $\mathcal{M}$, with respect to the inequalities we include.
We have $m_{\ell}+m_{\ell-1}\geq o_{\ell-1}$, as otherwise 
there would be a weight-augmenting 1-path; then we have 
$m_{\ell}+m_{\ell-1}+m_{\ell-2}\geq o_{\ell-2}+o_{\ell-1}$ so on so forth.
We also observe that each $o_j$ appears in $k-1$ inequalities. The first inequality containing $o_j$
has $m_{j+1}$ as the last term, and the last inequality containing $o_j$
has $m_j$ as the first term, as to include $o_j$ in $\mathcal{M}$, one has to remove $m_j$ and $m_{j+1}$ from $\mathcal{M}$.
 All together there are $k-1$ inequalities containing $m_{j}$ and $m_{j+1}$ together.

We collect the observations made so far below:
\begin{itemize}
\item $m_1$ appears in $(k-1)$ inequalities, as the first one is $m_1+m_2\geq o_1$,
the last one is $m_1+m_2+\cdots +m_k\geq o_1+ o_2+\cdots+ o_{k-1},$ and each time we add one more $m_i$.
\item $m_\ell$ appears in $(k-1)$ inequalities, as the first one is 
$m_{\ell-k+1}+\cdots+m_{\ell} \geq  o_{\ell-k+1}+\cdots+o_{\ell-1}$;
the last one is $m_{\ell-1}+m_{\ell} \geq  o_{\ell-1}$, and each time we remove one $m_i$;
\item all other $m_i$ appear in $k$ inequalities. For $i\leq k$, 
$m_i$ appears $k-i+1$ times for inequalities whose first term is $m_1$, then  
$i-1$ more times until $m_{i+1}$ becomes the first term.
For $i>k$, $m_i$ appears in all inequalities with $m_j$ as the first term
where $i-k+1\leq j\leq i$;
\item each $o_j$ appears in $k-1$ inequalities.
\end{itemize}
Under the light of these observations, by adding the listed equalities side-by-side we 
obtain
$$(k-1) m_1+(k-1)m_\ell + \sum _{i=2}^{\ell-1}k m_i \geq \sum_{j=1}^{\ell-1}(k-1) o_j\;,$$
therefore
$$k \sum _{i=1}^{\ell}m_i \geq (k-1)\sum_{j=1}^{\ell-1} o_j\;,$$
hence  $$w(\mathcal{M}\cap C)\geq \frac{k-1}{k}w(\mathcal{M}^*\cap C)$$
in this component.

Let us continue with showing the bound for the cycles with $2(\ell-1)$ edges as
shown in the third component of Figure~\ref{fig:mopluso}.
Let us take the part of the component with $m_1$ to $m_k$,
thus containing also $o_1$ to $o_{k-1}$.
We have
$$m_1+m_2+\cdots +m_k\geq o_1+ o_2+\cdots+ o_{k-1}\;,$$
otherwise there would be a weight-augmenting $(k-1)$-path.
We shift both sides of the inequality by one edge at a time and obtain 
$\ell-2$ more inequalities of the form
$$m_i+m_{i+1}+\cdots +m_{i+k-1}\geq o_i+\cdots+ o_{i+k-2}\;,$$
where an index $p>\ell-1$ is converted to $p-\ell+1$ and starts from 1.
This way we collect $\ell-1$ inequalities, where 
each $m_i$ appears in total $k$ times,
and each $o_j$ appears in total $k-1$ times.
Adding the $\ell-1$ inequalities side-by-side gives the 
desired result
$$w(\mathcal{M}\cap C)\geq \frac{k-1}{k}w(\mathcal{M}^*\cap C)$$
in a cyclic component $C$ with $2(\ell-1)$ edges where $\ell> k$.

We now look at a component $C$ of the second type 
shown in Figure~\ref{fig:mopluso}.
Let us add an artificial vertex $v$, an edge of zero weight from $v$ to the left vertex
of $o_{\ell-1}$, and add that edge to $\mathcal{M}$ to obtain a matching $\mathcal{M}'$ and a component $C'$.
\begin{claim}
Any weight-augmenting $k$-path for $\mathcal{M}'$ in $C'$ is also a weight-augmenting $k$-path for $\mathcal{M}$ in $C$.
\end{claim}
\textit{Proof of the claim:} If a weight-augmenting path does not contain the new edge we are done.
If it contains the new edge, then simply dropping that edge is still weight augmenting with the same
number of edges outside $\mathcal{M}$. $\square$

Based on the claim,
the component $C'$ is thus of the first form with $\ell-1$ edges outside $\mathcal{M}$. 
As the weights in $C$ are preserved in $C'$, we have the desired result
$$w(\mathcal{M}\cap C)\geq \frac{k-1}{k}w(\mathcal{M}^*\cap C)$$
for any component $C$ of the second type shown in Figure~\ref{fig:mopluso}.

For the components of the remaining form, the right-most one in Figure~\ref{fig:mopluso}, 
one can show the desired property by 
adding an edge $m_0$ of zero weight and converting the case the second one (and hence to the first one).

Since we have shown that in any component $C$ of $\mathcal{M}\oplus \mathcal{M}^*$ with at least $k$ edges from $\mathcal{M}^*$
we have $w(\mathcal{M}\cap C)\geq \frac{k-1}{k}w(\mathcal{M}^*\cap C)$, and there is no weight-augmenting $j$-path with $j < k$,
the claim is proved.
\end{proof}

\subsection{Proof of Theorem~\ref{thm:dyn-approx-ratio}}

\begin{proof}
The proof is adapted from the proof contained in~\cite{DBLP:conf/esa/Henzinger0P020} for the dynamic unweighted matching case.
Assume that no weight-augmenting $k$-path (\ie no augmenting path of length $\leq 2/\epsilon +3 $) exists.
Then, according to \Cref{thm:kappx}, the matching is a $(\frac{1/\epsilon + 1}{1/\epsilon}) = (1+\epsilon)$-approximate maximum matching. To see this, rewrite the length of the path to $2(1/\epsilon +1) +1$ and set $k=1/\epsilon+1$ in \Cref{thm:kappx}.  If there is such a path, then the probability of finding it is $\geq (\frac{1}{\Delta})^{2/\epsilon + 3}$:
after all, one possibility is that the random walker makes the ``correct'' decision at every vertex of the path. The probability that $\lambda$ random walks of length $2/\epsilon + 3$ do not find an augmenting path of length $2/\epsilon+3$ is $\leq (1-\frac{1}{\Delta^{2/\epsilon +3}})^\lambda \leq e^{-\frac{1}{\Delta^{2/\epsilon+3}}\cdot \lambda}$. Thus, for $\lambda \geq \Delta^{2/\epsilon+3} \log n$, the probability is $\leq 1/n$. 
\end{proof}

\clearpage
\newpage
\end{document}